\documentclass[showpacs,epsfig,preprintnumbers,amsmath,amssymb,nofootinbib,20pt]{revtex4-1}
\pdfoutput=1
\usepackage{graphicx}% Include figure files
\usepackage{dcolumn}% Align table columns on decimal point
\usepackage{bm}% bold math

\begin{document}
\def\d{{\rm d}}
%%%%%%%%%%%%%%%%%%%%%%%
\def\Epos{E_{\rm pos}}
\def\ap{\approx}
\def\eff{{\rm eft}}
\def\L{{\cal L}}
\newcommand{\vev}[1]{\langle {#1}\rangle}
\newcommand{\CL}   {C.L.}
\newcommand{\dof}  {d.o.f.}
\newcommand{\eVq}  {\text{EA}^2}
\newcommand{\Sol}  {\textsc{sol}}
\newcommand{\SlKm} {\textsc{sol+kam}}
\newcommand{\Atm}  {\textsc{atm}}
\newcommand{\Chooz}{\textsc{chooz}}
\newcommand{\Dms}  {\Delta m^2_\Sol}
\newcommand{\Dma}  {\Delta m^2_\Atm}
\newcommand{\Dcq}  {\Delta\chi^2}
\newcommand{\nbb}{$\beta\beta_{0\nu}$ }
\def\VEV#1{\left\langle #1\right\rangle}
\let\vev\VEV
\def\e6{E(6)}
\def\10{SO(10)}
\def\21{SA(2) $\otimes$ U(1) }
\def\321{$\mathrm{SU(3) \otimes SU(2) \otimes U(1)}$ }
\def\lr{SA(2)$_L \otimes$ SA(2)$_R \otimes$ U(1)}
\def\422{SA(4) $\otimes$ SA(2) $\otimes$ SA(2)}
\newcommand{\AHEP}{%
School of physics, Institute for Research in Fundamental Sciences
(IPM)\\P.O.Box 19395-5531, Tehran, Iran\\

  }
\newcommand{\Tehran}{%
School of physics, Institute for Research in Fundamental Sciences (IPM)
\\
P.O.Box 19395-5531, Tehran, Iran}
\def\roughly#1{\mathrel{\raise.3ex\hbox{$#1$\kern-.75em
      \lower1ex\hbox{$\sim$}}}} \def\lsim{\roughly<}
\def\gsim{\roughly>}
\def\ltap{\raisebox{-.4ex}{\rlap{$\sim$}} \raisebox{.4ex}{$<$}}
\def\gtap{\raisebox{-.4ex}{\rlap{$\sim$}} \raisebox{.4ex}{$>$}}
\def\lsim{\raise0.3ex\hbox{$\;<$\kern-0.75em\raise-1.1ex\hbox{$\sim\;$}}}
\def\gsim{\raise0.3ex\hbox{$\;>$\kern-0.75em\raise-1.1ex\hbox{$\sim\;$}}}

%\preprint{ IPM/P-2012}

%---------------------------------------------------

\title{Constraining Super-light Sterile Neutrino Scenario by JUNO and RENO-50 }

%----------------------------------------------------
\date{\today}
\author{P. Bakhti}\email{pouya\_bakhti@ipm.ir}
\author{Y. Farzan}\email{yasaman@theory.ipm.ac.ir}
%\author{A. Yu Smirnov}
\affiliation{\Tehran}
%\affiliation{\AHEP}
%--------------------------------------------------
\begin{abstract}
The Super-light Sterile Neutrino Scenario (SSNS) has been proposed
in the literature to explain the suppression of the upturn in the
low energy solar data. In this scenario, the mass splitting between
 the new mass eigenstate, $\nu_0$ and the standard $\nu_1$ is of
order of $\Delta m_{01}^2\sim 10^{-5}$ eV$^2$. Reactor neutrino
experiments with baseline larger than $\sim$20~km can help us to
probe this scenario. We study the potential of upcoming JUNO and
RENO-50 reactor experiments for discovering the superlight sterile
neutrino or constraining its mixing parameters. We study the
dependence of sensitivity to the SNSS  and find that the proposed
JUNO setup is very close to the optimal setup for  probing the SSNS.
\end{abstract}
%--------------------------------------------------------
 %\pacs{12.60.-i,14.80.-j,95.35.+d,98.70.Sa}
{\keywords{Neutrino, Leptonic CP Violation, Leptonic Unitary
Triangle, Beta Beam}}
%--------------------------------------------------------
\date{\today}
\maketitle
\section{Introduction}

 Three neutrino mass and mixing scheme has
been established as the solution to the lepton flavor violation
observed in solar, atmospheric, reactor and long baseline
accelerator neutrino data. Within this scheme, the neutrino
oscillation depends on two mass splittings ($\Delta m_{21}^2$ and
$\Delta m_{31}^2$), three mixing angles ($\theta_{12},\theta_{13}$
and $\theta_{23}$) and a CP-violating Dirac phase, $\delta_D$.
Throughout this letter, we will  use the standard parametrization in
\cite{pdg} for the three neutrino scheme. All the neutrino
parameters entering neutrino oscillation formula have been measured
except for the value of $\delta_D$ and sgn$(\Delta m_{31}^2)$. To
determine these last two parameters, studies and various plans are
underway.

 There are however some anomalies that
cannot be explained within the present three neutrino scheme. The
most famous examples are  the $\nu_\mu\to \nu_e$ and/or
$\bar\nu_\mu\to \bar\nu_e$ transitions in short baseline LSND and
MiniBooNE experiments \cite{LSND},   reactor neutrino deficit
\cite{rec-an} and Gallium anomaly \cite{kol}. To explain these
anomalies, various scenarios have been suggested among which
arguably the most popular one is adding one or more sterile
neutrino(s) with mass splitting of order of O(1) eV$^2$ with the
active ones. With ICECUBE atmospheric neutrino data, this solution
can be tested \cite{Esmaili:2013vza}. For a comprehensive study of
sterile neutrinos and model building issues, see Refs.
\cite{white,keV}.

Within the standard LMA MSW solution to the solar puzzle, an upturn
of the energy spectrum of events at energies below 8 MeV is
expected. However, data from SNO, Super-Kamiokande and Borexino
experiments does not display such an upturn. In
\cite{Holanda1,Holanda2}, a scenario is proposed to explain the
suppression of the upturn. The scenario is based on adding a new
sterile neutrino which weakly mixes with the active neutrinos. The
new mass eigenstate is called $\nu_0$ and its mass is denoted by
$m_0$. To explain the suppression of the upturn in the low energy
solar data, it is shown that $\Delta m_{01}^2 \sim (0.7-2)\cdot
10^{-5}$~eV$^2$ and $\sin^2 2 \alpha \sim 10^{-3}$ where $\alpha$
determines the $\nu_s$ mixing in $\nu_1$ and/or $\nu_2$. We refer to
this scenario as SSNS standing for Superlight Sterile Neutrino
Scenario. Notice however that the low energy solar neutrino data
still suffers from large uncertainties and the suppression of the
upturn is far from being established. Borexino, KamLAND solar
\cite{Kam} and SNO+ \cite{SNO} will reduce the uncertainty.
Ref.~\cite{Dev} demonstrates that the SSNS can be embedded within a
concrete model.

Notice that in this scenario, $\Delta m_{01}^2 \sim \Delta
m_{21}^2$. As a result, to test this scenario a long baseline setup
such as the ones used to determine mass ordering can be employed.
Now that a relatively large value for $\theta_{13}$ is established
by reactor experiments \cite{An:2012eh}, new phases of these
experiments are being studied to help  determining the mass
hierarchy. JUNO (Jiangmen Underground Neutrino Observatory formerly
known as Daya BAY II)  \cite{Learned} and Reno-50 \cite{reno50}
experiments are specifically proposed for this purpose but can also
be used for other measurements such as precision measurement of
$\theta_{12}$ or studies of geoneutrinos \cite{Learned}. The
baseline in these experiments are about 50 km, leading to $\Delta
m_{01}^2 L/(2 E_\nu)\sim 0.4 (\Delta m_{01}^2/10^{-5}{\rm eV}^2)
(L/50~{\rm km})(3~{\rm MeV}/ E_\nu)$ so this setup is suitable to
search for the effects of a sterile neutrino with $\Delta
m_{01}^2\ll 1 ~{\rm eV}^2$. In this letter, we  examine the ability
of such long baseline reactor experiments in testing this scenario.
We then study the dependence of the sensitivity to SSNS on baseline
to search for optimal setup to test the SSNS.

 In section II, we describe the four neutrino scenario in which we are
  interested and outline its new parameters. We then summarize  the bounds that
   already exist on the parameters.
In section III,  we describe the setups and the inputs for numerucal
analysis. In section IV, we study oscillation pattern for SSNS and
present our results. In section V, we summarize our conclusions.
%%%%%%%%%%%%%%%%%%%%%%%%%%%%%%%%%%%%%%%%% for
\section{Four neutrino scheme }
%%%%%%%%%%%%%%%%%%%%%%%%%%%%%%%%%%%%%%%
The four neutrino mixing matrix can be described by six mixing
angles and three physical Dirac phases. If the neutrinos are of
Majorana type, there will  also be three Majorana CP-violating
phases which do not show up in the neutrino oscillation patterns.
Following the notation in \cite{Holanda2}, we call the mass
eigenstates as $(\nu_0,\nu_1,\nu_2,\nu_3)$ with mass eigenvalues
$(m_0,m_1,m_2,m_3)$. The flavor eigenstates are related to mass
eigenstates by a $4\times 4$ unitary matrix, $U$ as follows
\begin{eqnarray}
\left( \begin{matrix} \nu_s \cr \nu_e \cr \nu_\mu \cr \nu_\tau
\end{matrix} \right)=U\cdot \left( \begin{matrix} \nu_0 \cr \nu_1 \cr \nu_2 \cr
\nu_3
\end{matrix} \right)
\end{eqnarray}
Notice that $\nu_0$ is not necessarily the lightest state. In fact,
Ref. \cite{Holanda2} has shown that suppression of the upturn in low
energy solar neutrino data  implies $m_1<m_0<m_2$.

There is a standard way of parameterizing the $3\times 3$ PMNS
matrix \cite{pdg} but there is not such a standard form for the
parametrization of the $4 \times 4$ mixing matrix. To be compatible
with the notation in \cite{Holanda2}, we use the following
parametrization:
\begin{eqnarray}
U\equiv\left( \begin{matrix}1 & 0\cr 0 & U_{PMNS}\end{matrix}
\right)\cdot U_S
\end{eqnarray}
where $U_{PMNS}$ is the standard $3\times 3$ PMNS matrix and $U_S$
is the matrix mixing the sterile neutrino with active ones:
\begin{eqnarray} \label{US}
U_S=\left( \begin{matrix} \cos \alpha  & \sin\alpha e^{i \delta_1} &
0 & 0\cr -\sin \alpha e^{-i \delta_1} &   \cos \alpha&0 & 0 \cr 0
 & 0 & 1 & 0 \cr 0 & 0 & 0 & 1\end{matrix} \right).
 \left( \begin{matrix} \cos \gamma  & 0 & \sin\gamma   & 0\cr 0 &   1 &0 & 0 \cr
-\sin\gamma
 & 0 & \cos\gamma  & 0 \cr 0 & 0 & 0 & 1\end{matrix} \right).
 \left( \begin{matrix} \cos \beta  &  0 & 0 & \sin\beta e^{i \delta_2} \cr 0 &  1 &0 & 0 \cr 0
 & 0 & 1 & 0 \cr - \sin\beta e^{-i \delta_2}& 0 & 0 & \cos \beta \end{matrix} \right)\ .\end{eqnarray}
 To avoid the  current bounds we take the mixing between $\nu_s$ and active
neutrinos to be small.
 Notice that $\alpha$ and $\beta$ can be identified with the same  parameters in
 \cite{Holanda2}. In the parameter range of our interest where
 $\Delta m_{01}^2, \Delta m_{20}^2, \Delta m_{21}^2 \ll | \Delta
 m_{31}^2|$, for intermediate baselines for which $ \Delta
 m_{31}^2 L/E\sim \pi$ (and therefore $ \Delta
 m_{01}^2 L/E\ll \pi$), $\alpha$ and $\gamma$ parameters cannot be
 resolved. However, such experiments are sensitive to  $\beta$. The atmospheric data  \cite{atm} and long baseline experiment
 MINOS \cite{MINOS}
 already constrain \begin{equation} \label{MIN-at}\sin^2 \beta
<0.2.\end{equation}
% MINOS+ will be able to
% further constrain $\beta$ \cite{MINOS+}
A stronger bound comes from cosmology. Recent PLANCK data constrains
the effective number of relativistic degrees of freedom at the
recombination era, $N_{eff}$ \cite{Planck}. If $\beta$ or $\gamma$
are  large enough, $\nu_s$ can reach thermal equilibrium at the
early universe and can be considered as an extra degree of freedom,
contributing $\Delta N_{eff}=1$. From this observation, Ref.
\cite{Mirizzi} finds $\sin^2\beta, \sin^2 \gamma<10^{-3}$ ({\it see}
also, \cite{atm,cosmo}). One should however remember that there is a
deviation between the values of Hubble constant derived from the
PLANCK data and from the local measurements. This discrepancy can be
settled by an extra degree of freedom \cite{1X} so the bounds found
in Ref. \cite{Mirizzi} can be relaxed.

%%%%%%%%%%%%%%%%%%%%%%%%%%%%%%
%%%%%%%%%%%%%%%%%%%%%%%%%%%%%%%%%%%%
\section{ JUNO and RENO-50 experiments}
%%%%%%%%%%%%%%%%%%%%%%%%%%%%%%%%%
%%%%%%%%%%%%%%%%%%%%%%%%%%%%%%%%
The potential of reactor neutrino experiments with a baseline of
$\sim $ 50 km  for determining  the neutrino mass ordering has been
extensively studied in the literature
\cite{piai,Blen,Evslin,Li,Ciuffoli}. The main goal of JUNO and
RENO-50 experiments, which according to schedules \cite{reno50} will
be ready for data taking in 2020, is determining the sign of $\Delta
m_{31}^2$. For this purpose as we explain in the next section, high
energy resolution is required. Both detectors employ liquid
scintillator and aim at an energy resolution  of $3\%
/\sqrt{E_\nu/{\rm MeV}}$. It is shown that in order to determine
sgn($\Delta m_{31}^2$), the difference between the distances of
different reactors contributing to the flux of the detector should
be less than 500 meters \cite{Li}. Considering this restriction, the
best location for JUNO is found to be at a 52 km distance from
Xangjiang and Taishan reactor complexes \cite{Li}. The Daya Bay
reactor, which is the main contributor to the flux of the current
Daya Bay I experiment, is located at a distance of 215 km and the
planned Huizhou reactor will be 265 km away. The combined power of
the two close reactor complexes is 36 GW. The flux from the far
reactors is a troublesome background for the purpose of determining
the hierarchy; however, as we shall discuss in the next section for
probing the SSNS, the flux from the far sources will also be
helpful. The JUNO setup will not have a near detector so the flux
normalization uncertainty will be around 3 \%. As we shall discuss
in the next section this uncertainty will not be a problem for
probing the SSNS. For our analysis, we use the baselines and powers
listed in table 1 of \cite{Li} for JUNO.

The RENO-50 setup uses the same reactor used for the current RENO
experiment. The baseline for RENO-50 will be 47 km. The total power
is 16.4 GW. The near detector can be used to measure the total flux
so the flux normalization uncertainty can be reduced to 0.3\%
\cite{Seo}. In principle, the KamLAND data could be also used for
testing the SSNS. While the total $\nu$ flux at KamLAND is
comparable to that in JUNO or RENO-50, the new detectors are larger
by one order of magnitude. We therefore focus on the JUNO and
RENO-50 experiments and do not consider the KamLAND results.

We use the GLoBES software \cite{Globes} implementing four neutrino
scheme \cite{Kopp:2007ne} for our analysis. We take 20~kton
(18~kton) scintillator detector for JUNO (RENO-50) experiment. We
assume energy resolution equal to $3\%/\sqrt{E(MeV)}$, and 62 bins
are considered from 1.8 MeV to 8 MeV. We employ the energy spectrum
for the  reactor neutrino as shown in
\cite{Murayama:2000iq,Eguchi:2002dm}. The cross section of neutrinos
is taken from \cite{Vogel:1999zy}. We take the  uncertainties  as
well as the best fit values    of the parameters of the
three-neutrino scheme from \cite{GonzalezGarcia:2012sz}.  We use the
pull-method to account for the uncertainties in the parameters of
the three neutrino mass scheme. We take the ordering of the mass
eigenstates to be normal and set $\delta_D=0$. According to
\cite{Learned}, the main sources of background are (i) accidental
background; (ii) $^{13} C(\alpha,n) ^{16}O$ background and (iii)
Geoneutrino background. We take the spectrum of these sources of
background from \cite{bs} and normalize each background flux as
described in \cite{Learned}.

%%%%%%%%%%%%%%%%%%%%%%%%%%%%%%%%%%
%%%%%%%%%%%%%%%%%%%%%%%%%%%%%%%%%5
 \section{Oscillation of neutrinos}
%%%%%%%%%%%%%%%%%%%%%%%%%%%%%%%%
%%%%%%%%%%%%%%%%%%%%%%%%%%%%%%
 The energy of the reactor
 neutrinos is of order of a few MeV. At such energies matter effects
 on the oscillation probability are negligible: $V_{eff}\sim
  G_F n_e \sim G_F n_n\ll \Delta m_{01}^2/E_\nu <\Delta
  m_{21}^2/E_\nu\ll |\Delta m_{31}^2/E_\nu|$. The oscillation
  probability can be written as
  \begin{equation} \label{PeeBB}
  P(\bar{\nu}_e \to \bar{\nu}_e)
=\left|M_0e^{i \Delta_0}+M_1e^{i \Delta_1}+M_2e^{i \Delta_2}+M_3e^{i
\Delta_3}\right|^2
\end{equation}
where $\Delta_i =m_i^2L/2E_\nu$ and \begin{eqnarray}
\label{M0-3}M_0&=&|\cos \beta (-e^{-i \delta_1}\cos \gamma \cos
\theta_{12} \cos \theta_{13}\sin \alpha - \cos \theta_{13}\sin
\gamma \sin \theta_{12})-e^{-i(\delta_D+\delta_2)}\sin \beta \sin
\theta_{13}|^2, \cr M_1&=&|\cos \alpha \cos
\theta_{12}\cos\theta_{13}|^2 \cr
M_2&=&|-e^{-i\delta_1}\cos\theta_{12}\cos\theta_{13}\sin \alpha \sin
\gamma+\cos\gamma \cos \theta_{13}\sin \theta_{12}|^2 \cr M_3 &=&
|\sin \beta (-e^{-i\delta_1}\cos \gamma \cos \theta_{12}\cos
\theta_{13}\sin \alpha -\cos \theta_{13}\sin \gamma \sin
\theta_{12})+e^{-i(\delta_D+\delta_2)} \cos \beta \sin
\theta_{13}|^2.
\end{eqnarray} In the absence of mixing with the sterile neutrinos,
we recover the standard formula: \begin{equation} \label{os}
P(\bar{\nu}_e \to \bar{\nu}_e)
=\left|\cos^2\theta_{13}\cos^2\theta_{12}e^{i
\Delta_1}+\cos^2\theta_{13} \sin^2\theta_{12}e^{i \Delta_2}+\sin^2
\theta_{13}e^{i \Delta_3}\right|^2\end{equation} The interference of
the second and third terms in the oscillation formula (given by
$\cos(\Delta_3-\Delta_2)$) is sensitive to sgn$(\Delta m_{31}^2)$.
To determine the hierarchy, the oscillatory mode given by $\Delta
m_{32}^2$ has to be resolved which  requires excellent  energy
resolution \cite{res}.
%, $\delta E/E$, better than $0.6~\%~ (E_\nu/
%4 ~{\rm MeV})~(50~{\rm km}/L)~(2.4\times 10^{-3}~{\rm eV}^2/|\Delta
%m_{32}^2|)$ and enough statistics in energy bins with size below
%0.01 MeV.
Moreover, to resolve this mode, the difference between the
distances of different sources should be smaller than the
oscillation length associated with $\Delta m_{31}^2$; otherwise,
averaging effects will wash out the sensitivity to the sign. In Eq.
(\ref{os}), there are four oscillatory modes (proportional to  $\cos
(\Delta_1-\Delta_2)$, $\cos (\Delta_1-\Delta_3)$, $\cos
(\Delta_3-\Delta_2)$ and constant) which are in principle enough to
determine all the parameters entering Eq. (\ref{os}), even without
the knowledge of the total normalization of the flux. Notice that
the CP-violating phase $\delta_D$ and mixing angle $\theta_{23}$ do
not enter Eq. (\ref{os}) so unlike the case of long baseline
experiments, here the derivation of sign($ \Delta m_{31}^2$) will
not suffer from ambiguity induced by the octant or $\delta_D$
degeneracy. The matter effects will cause a deviation of effective
mixing angles and mass splittings from the vacuum values of order of
1 \%. For the purpose of the precision measurement of $\theta_{12}$
and $\Delta m_{21}^2$, matter effects have to be taken into account
but for our analysis, we can safely neglect the matter
effects. %However, the GLoBES software \cite{Globes} that we use for
% our numerical calculation takes into account the matter effects.

Let us now discuss the effects of each mixing by the sterile
neutrino one by one. In the end, we will discuss the case when all
mixings are nonzero.

\begin{figure}[htp]

 \vspace{-5 cm} \hspace{+7mm}

    \includegraphics[bb=0 0 800 1000,scale=0.75]{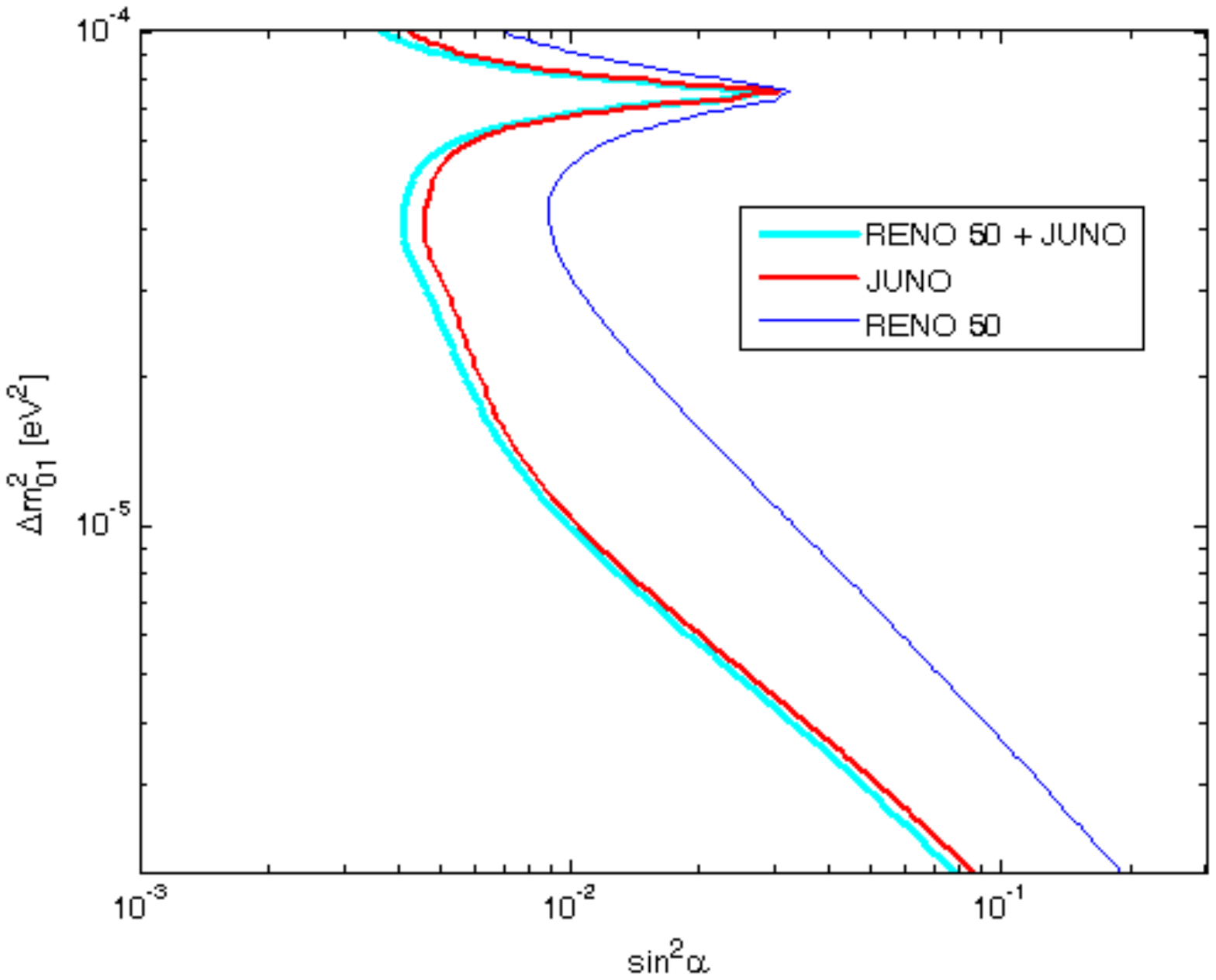}

 \caption{The 95\% C.L. upper bound on $\sin^2 \alpha$ versus
$\Delta m_{01}^2$.  Five years of data taking with JUNO and RENO-50
experiments are assumed. The true values of $\alpha$, $\gamma$ and
$\beta$ are set to zero. The  blue (thin), red (intermediate) and
cyan (thick) curves correspond respectively to the RENO-50, JUNO and
combined JUNO and RENO-50 performances. The values of the parameters
of the three neutrino mass scheme are set to the best values found
in Ref. \cite{GonzalezGarcia:2012sz} and the same uncertainties are
assumed. We have taken the ordering of the mass eigenstates to be
normal and have set $\delta_D=0$.
  \label{alpha-bound} }
\end{figure}

\begin{figure}[htp]

  \vspace{-5 cm} \hspace{+7mm}

    \includegraphics[bb=0 0 800 1000,scale=0.75]{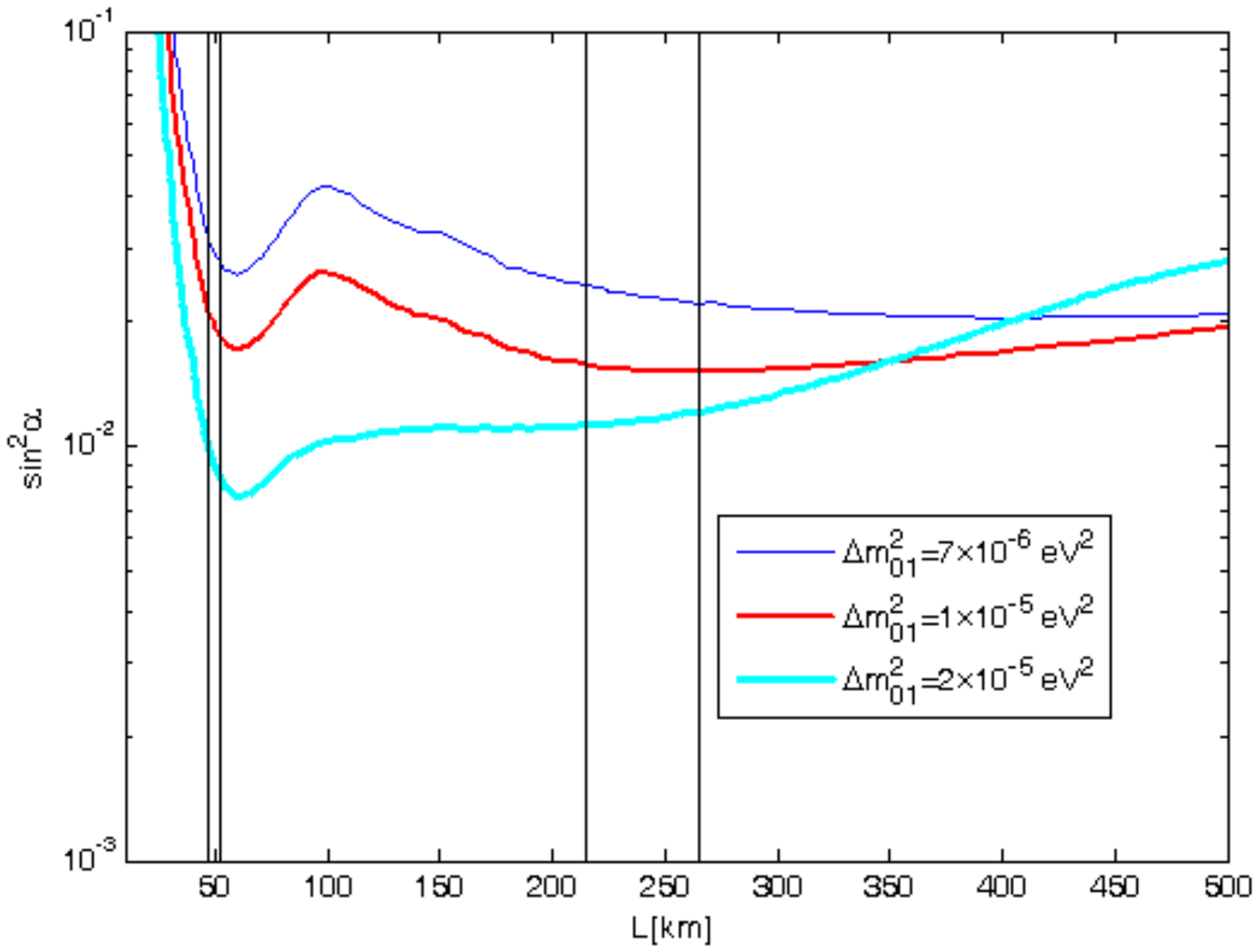}

 \caption{The 95\% C.L. upper bound on $\sin^2 \alpha$ versus
the baseline.  Five years of data taking with a 20 kton detector and
36 GW reactor source are assumed. The true values of $\alpha$,
$\gamma$ and $\beta$ are set to zero. The  cyan (thick), red
(intermediate) and blue (thin) curves correspond respectively to
$2\times 10^{-5}$ eV$^2$,  $ 10^{-5}$ eV$^2$ and $0.7\times 10^{-5}$
eV$^2$. The values   of the parameters of the three neutrino mass
scheme are set to the best values found in Ref.
\cite{GonzalezGarcia:2012sz}  and the same uncertainties are
assumed.  We have taken the ordering of the mass eigenstates to be
normal and have set $\delta_D=0$. From left to right, the vertical
lines show the baseline of RENO-50, the distance between the JUNO
detector to the Taishan or Xangjiang reactors, the distance between
the JUNO detector to the Daya Bay reactor and the distance between
the JUNO detector to the Huizhou reactor.
  \label{alpha-L} }
\end{figure}

\begin{figure}[htp]

 \vspace{-5 cm} \hspace{+7mm}

    \includegraphics[bb=0 0 800 1000,scale=0.75]{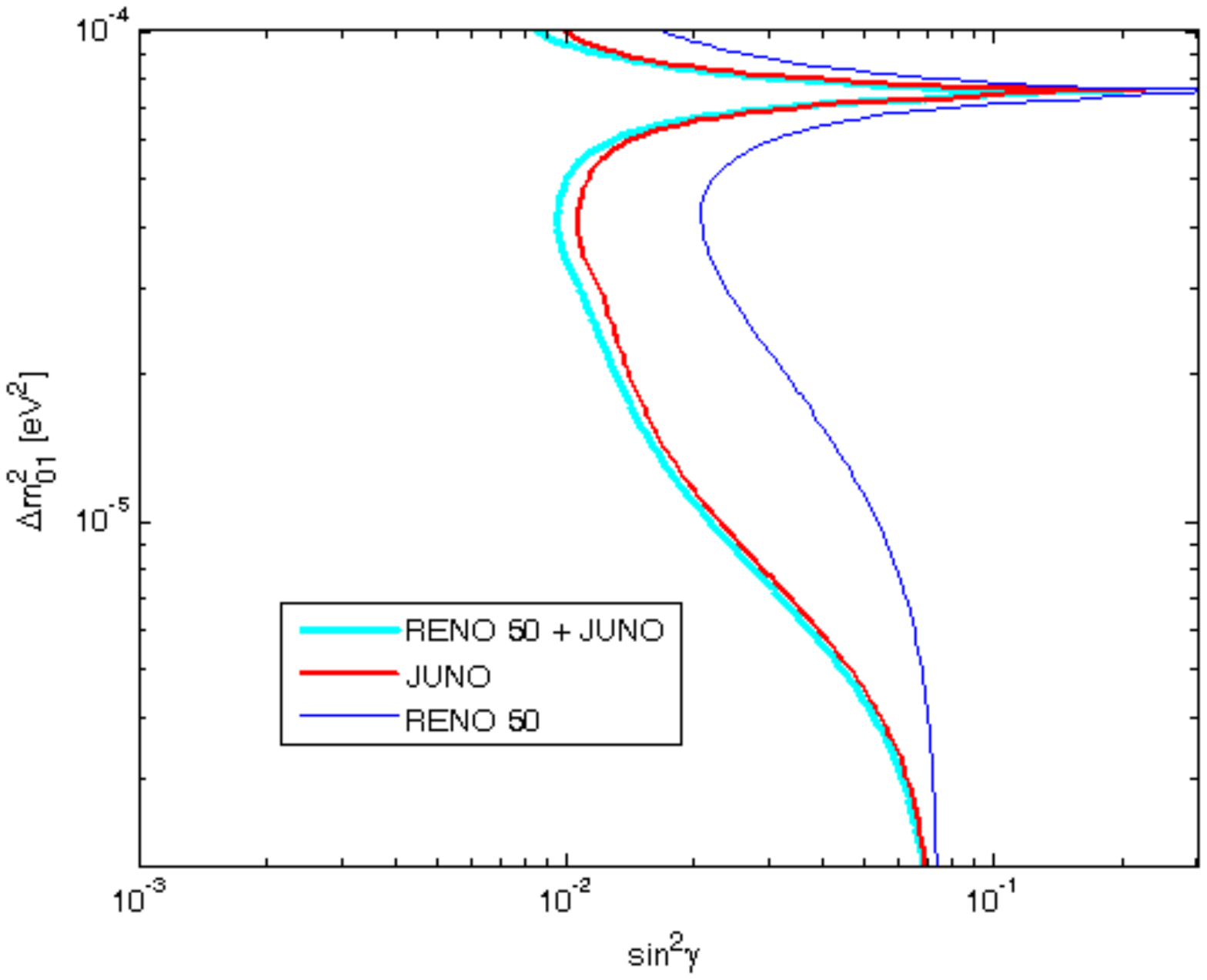}

 \caption{ The 95\% C.L. upper bound on $\sin^2 \gamma$ versus
$\Delta m_{01}^2$. The rest of the description is similar to that of
Fig. \ref{alpha-bound}. \label{gamma-bound}}
\end{figure}

\begin{figure}[htp]

\vspace{-5 cm} \hspace{+7mm}

    \includegraphics[bb=0 0 800 1000,scale=0.75]{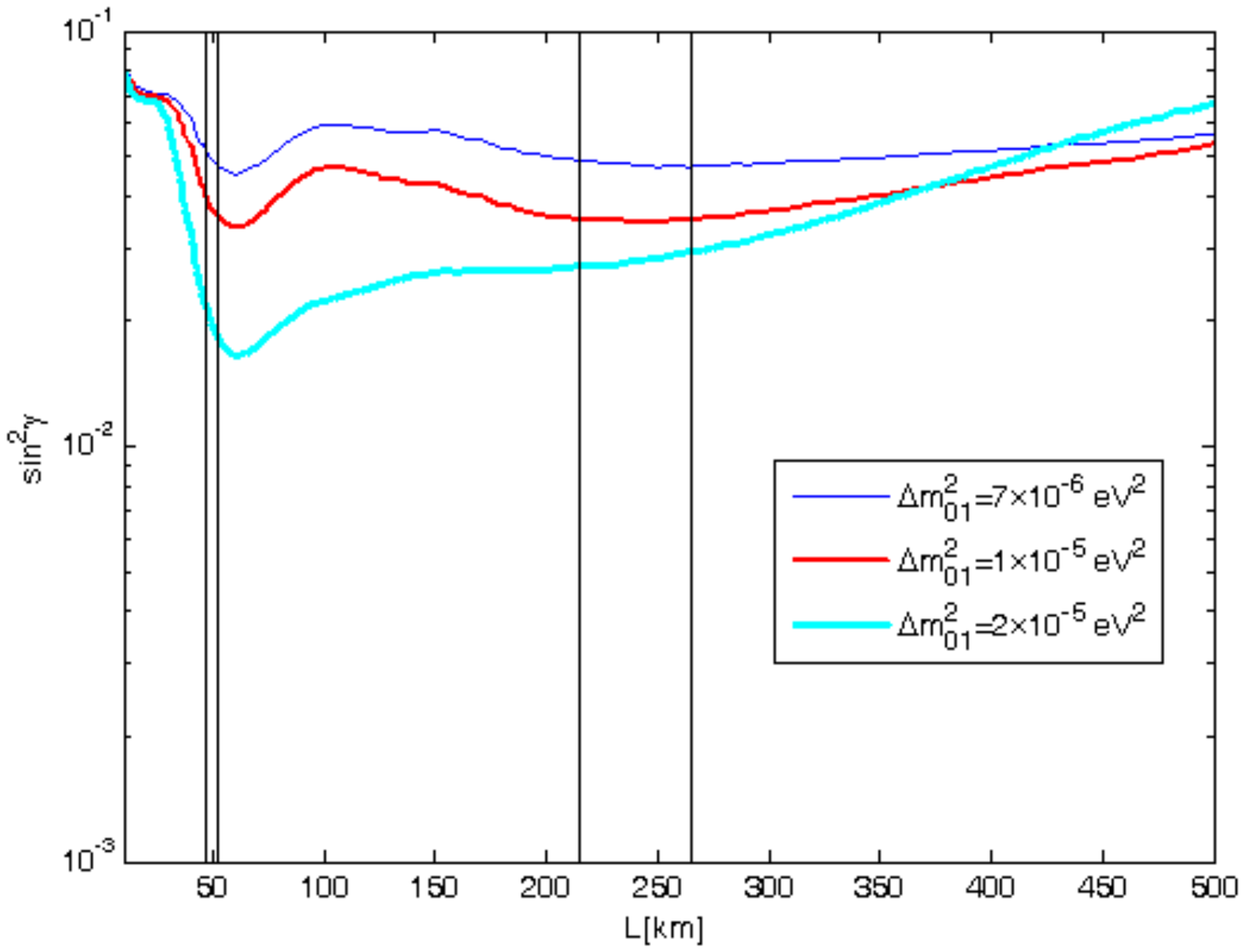}

 \caption{ The 95\% C.L. upper bound on $\sin^2 \gamma$ versus
the baseline. The rest of the description is similar to that of Fig.
\ref{alpha-L}.\label{gamma-L}  }
\end{figure}

\begin{enumerate}\item {\it Case I, $\sin \beta=\sin \gamma=0$ and $\alpha \ne 0$}:
In this limit, the oscillation probability can be written as
\begin{equation} \label{os-a}
P(\bar{\nu}_e \to \bar{\nu}_e)
=\left|\cos^2\theta_{13}\cos^2\theta_{12}\sin^2\alpha e^{i
\Delta_0}+\cos^2\alpha \cos^2\theta_{13}\cos^2\theta_{12}e^{i
\Delta_1}+\cos^2\theta_{13} \sin^2\theta_{12}e^{i \Delta_2}+\sin^2
\theta_{13}e^{i \Delta_3}\right|^2\end{equation}

By measuring the frequency and amplitude of the oscillatory modes
$\cos(\Delta_0-\Delta_1)$ and/or $\cos(\Delta_0-\Delta_2)$, the
values of $\alpha$ and $\Delta m_{01}^2$ can be determined.  Notice
that to derive these parameters a moderate precision in knowledge of
the flux normalization as well as the energy reconstruction
resolution will be enough. In other words, the results will not be
sensitive to uncertainty in flux normalization or energy resolution.
Since  sgn($\Delta m_{31}^2$) and $(\alpha,\Delta m_{01}^2)$ are
determined by completely separate oscillatory modes, the presence of
the mixing $\alpha$ does not affect the determination of sgn($\Delta
m_{31}^2)$. As expected in the  $\Delta m_{01}^2\to 0$ limit, the
sensitivity to $\alpha$ is lost.

Fig. \ref{alpha-bound} shows the bound that JUNO and RENO-50
experiments with 5 years of data taking can impose on $\alpha$. To
draw this figure, we have set $\alpha=\beta=\gamma=0$  and have
varied the true value of $\Delta m_{01}^2$ using the GLoBES software
\cite{Globes}. For given $\Delta m_{01}^2$  the  value of $\sin^2
\alpha$ shown by curves can be distinguished from $\alpha=0$ at 95
\% C.L. The  blue (thin), red (intermediate) and  cyan (thick)
curves correspond respectively to the RENO-50, JUNO and combined
JUNO and RENO-50 results after 5 years of data taking. As expected
for $\Delta m_{01}^2<2\times 10^{-5}$ eV$^2$, the bound on $\sin^2
\alpha$ becomes weaker.  We also draw plots (not shown here)
displaying the minimum $\sin^2 \alpha$ that can be distinguished
from $\sin^2 \alpha=0$. That is we varied the ``true'' values of
$\Delta m_{01}^2$ and $\alpha$, plotting the value of $\sin^2\alpha$
versus $\Delta m_{01}^2$ for which $\sin^2\alpha=0$ can be
distinguished from it by 95 \% C.L. As expected the results were
only slightly different from those shown in Fig.~1.

Fig. \ref{alpha-bound} shows that as $\Delta m_{01}^2 \to \Delta
m_{21}^2$, the sensitivity to $\sin^2 \alpha$ becomes weaker.
Moreover at this point, the three curves converge. From Eq
(\ref{os-a}), it is straightforward to verify that as $\Delta_0 \to
\Delta_2$, the effect of $\alpha$ can be mimicked by a small shift
in $\theta_{12}$. Thus, in this limit the bound on $\sin^2 \alpha$
is determined by the uncertainty in $\theta_{12}$: $\sin^2 \alpha
<\delta \sin^2 \theta_{12}/\cos ^2 \theta_{12}$. Since the effects
of $\theta_{13}$ are already subdominant the uncertainty in
$\theta_{13}$ does not affect the sensitivity to $\alpha$ and other
new mixing parameters. We have checked this statement by turning off
the uncertainty of $\theta_{13}$ and $\theta_{12}$ one by one. While
the uncertainty in $\theta_{13}$ seems to be irrelevant, setting the
uncertainty in $\theta_{12}$ equal to zero, we lose the feature at
$\Delta m_{01}^2 \to \Delta m_{21}^2$.   The uncertainty of the
solar mass splitting ({\it i.e.,} $\delta (\Delta m_{21}^2))$
renders the experiment unable to constrain $\alpha$ for $\Delta
m_{01}^2<\delta (\Delta m_{21}^2)\sim 2 \times 10^{-6}$ eV$^2$.

 Fig.  \ref{alpha-L} shows the bound that can be derived on
$\sin^2 \alpha$ versus baseline for a reactor with 36 GW power and
20 kton detector. For any other values of power and the detector
size, the dependence on baseline should be similar but the curves
can be shifted up or down. It is noteworthy that there is a local
minimum close to 50 km (close to the baseline of JUNO as well as
that of RENO-50) which makes these setups close to ideal for the
purpose of probing the SSNS. For $\Delta m_{01}^2 \leq 10^{-5}$
eV$^2$, the performance of a setup with baseline farther than 200 km
is better. Thus, the Daya Bay or Huizhou reactors or other farther
reactor sources contributing to the flux at JUNO can be  even more
helpful than the closer sources for probing the SSNS. Considering
Eq.~(\ref{os-a}), this is understandable because to discern
$\sin^2\alpha$, $\Delta m_{01}^2L/E$ should be sizeable.
 \item
{\it Case II, $\sin \alpha=\sin \beta=0$ and $\gamma \ne 0$}: In
this limit,
\begin{equation} \label{os-g}
P(\bar{\nu}_e \to \bar{\nu}_e)
=\left|\cos^2\theta_{13}\sin^2\theta_{12}\sin^2\gamma e^{i
\Delta_0}+ \cos^2\theta_{13}\cos^2\theta_{12}e^{i
\Delta_1}+\cos^2\theta_{13} \sin^2\theta_{12}\cos^2 \gamma e^{i
\Delta_2}+\sin^2 \theta_{13}e^{i \Delta_3}\right|^2.\end{equation} A
discussion similar to case I holds here, too. The only difference is
that  for $\Delta m_{01}^2 \to 0$, the sensitivity to $\gamma$ is
limited by $\delta \sin^2 \theta_{12}/\sin^2 \theta_{12}$ but the
sensitivity is lost when $\Delta m_{01}^2 \to \Delta m_{21}^2 $
regardless of the uncertainty in $\sin^2 \theta_{12}$. Fig.
\ref{gamma-bound} displays the bound that JUNO and RENO-50
experiments with 5 years of data taking can impose on $\gamma$. As
expected at $\Delta m_{01}^2=\Delta m_{21}^2 $, the sensitivity to
$\sin \gamma$ is lost. The bound on $\sin^2 \alpha$ for $\Delta
m_{01}^2\to \Delta m_{21}^2 $ is about $\tan^2 \theta_{12}$ times
the bound on $\sin^2 \gamma$ for $\Delta m_{01}^2\ll \Delta m_{21}^2
$.
%That is because $\sin^2\alpha$ in Eq. (\ref{os-a}) is multiplied
%by $\cos^2 \theta_{12}$ but $\sin^2\gamma$ in Eq. (\ref{os-g}) is
%multiplied by $\sin^2 \theta_{12}$.

Fig.  \ref{gamma-L} displays the bound that can be imposed on $ \sin
^2 \gamma$ versus baseline for different values of $\Delta m_{01}^2$
favored by the solar data \cite{Holanda1,Holanda2}. Independently of
$\Delta m_{01}^2$, the optimal baseline is located close to 50 km
which means the JUNO and RENO-50 setups, despite being designed for
another purpose, are close to optimal setup for probing SNSS, too.
Unlike the case I, here there is no other minimum for smaller
$\Delta m_{01}^2$. That is because to discern $\sin^2\gamma$,
$\Delta m_{01}^2L/E$ need not to  be sizeable.
\item
{\it Case III, $\sin \alpha=\sin \gamma=0$ and $\beta \ne 0$}: In
this limit,
\begin{equation} \label{os-b}
P(\bar{\nu}_e \to \bar{\nu}_e) =\left|\sin^2\theta_{13}\sin^2\beta
e^{i \Delta_0}+ \cos^2\theta_{13}\cos^2\theta_{12}e^{i
\Delta_1}+\cos^2\theta_{13} \sin^2\theta_{12} e^{i \Delta_2}+\sin^2
\theta_{13}\cos^2 \beta e^{i \Delta_3}\right|^2\end{equation} By
deriving the amplitudes of the oscillatory modes given by
$\cos(\Delta_0-\Delta_1)$ and $\cos(\Delta_0-\Delta_2)$,  the value
of $\sin \beta$ can be extracted while the frequency of these modes
gives $\Delta m_{01}^2$. Like the case II,  sensitivity to $\sin
\beta$ persists even in the limit $\Delta m_{01}^2 \to 0$ so setups
with longer baselines have no advantage. Increasing the baseline,
the sensitivity to the SSNS deteriorates because of the flux
decrease. Similarly to the cases I and II, to derive $\sin \beta$
from these modes a moderate precision in the flux normalization and
a moderate energy resolution to reconstruct the modes given by
$\Delta m_{01}^2$ are sufficient. The amplitudes of the oscillatory
$\cos (\Delta_3-\Delta_1)$ and $\cos (\Delta_3-\Delta_2)$ terms here
are suppressed by $\cos^2 \beta$. For $\beta \ll 1$, this
suppression is negligible and sgn($\Delta m_{31}^2$) can be derived,
like in the standard case. From  Eq. (\ref{os-b}), we observe that
the sensitivity to $\sin^2 \beta$  is suppressed by a factor of
$\sin^2 \theta_{13}$. Thus, as shown in Fig. \ref{beta-bound} the
bound that can be imposed on $\sin^2 \beta$ by this method is
relatively weak; however, they are slightly stronger than the
present MINOS bound \cite{MINOS} shown in Eq. (\ref{MIN-at}).
\end{enumerate}
We have found that with 20 years of data taking with JUNO and
RENO-50 at $\Delta m_{01}^2=2\times 10^{-5}$ eV$^2$, the bound on
$\sin^2\alpha$ and $\sin^2\gamma$ can be at best lowered down to
$2.8\times 10^{-3}$ and $4.2\times 10^{-3}$, respectively. For
smaller values of $\Delta m_{01}^2$, the bound on $\sin^2\alpha$
will be weaker but the bound on $\sin^2\gamma$ will remain the same.
Thus, with this setup it is possible to only marginally touch the
parameter range indicated in \cite{Holanda1,Holanda2} ({\it i.e.,}
$\Delta m_{01}^2=(0.7-2)\times 10^{-5}$~eV$^2$ and
$\sin^2\alpha,\sin^2 \gamma\sim 10^{-3}$). By increasing the
baseline, $\Delta m_{01}^2$ can be resolved better. If the future
solar data finds more evidence in support of the SSNS, a setup with
multiple sources such as KamLAND but with  $\sim$ 20 kton size
detector will be a more suitable setup to test SSNS as, unlike the
case of determining sgn($\Delta m_{31}^2$), the contributions from
different sources add up to determine $\alpha$ and $\gamma$.

 The following
comments are in order:
\begin{itemize}
\item
As seen from Figures \ref{alpha-bound}, \ref{gamma-bound} and
\ref{beta-bound} the performance of JUNO is overally better than
RENO-50. The main reason is that the power of JUNO  is considerably
larger than that of RENO-50. Moreover as seen in Figs. \ref{alpha-L}
and \ref{gamma-L}, the main baseline of JUNO is close to the optimal
baseline.
\item
The background in detectors \cite{Learned} is low enough not to
affect probing the SSNS. We checked for the effects of the
background by turning it off  and found that the change in results
was negligible.
\item
 The interference of the first and third terms  in Eqs. (\ref{os-a},\ref{os-g},\ref{os-b}) as well as in Eq.
(\ref{PeeBB}), which is given by $$
\cos(\Delta_2-\Delta_0)=\cos(\Delta_2-\Delta_1)\cos(\Delta_1-\Delta_0)-\sin(\Delta_2-\Delta_1)\sin(\Delta_1-\Delta_0),$$
is already sensitive to the sign of $\Delta m_{01}^2$ even though
matter effects are suppressed.
\item Since the information on new
mixing angles come from the oscillating modes with a oscillation
length much larger than 1~km, unlike the case of deriving the
sign($\Delta m_{31}^2$), here the distance difference between the
different sources contributing to the flux is not a cause of
concern.
\item
In the above cases, none of the CP-violating phases appear in
$P(\bar{\nu}_e \to \bar{\nu}_e)$. However, when two or more mixing
angles are nonzero, the Dirac CP-violating phases show up in
$P(\bar{\nu}_e \to \bar{\nu}_e)$.
\item The dependence of
$P(\bar{\nu}_e \to \bar{\nu}_e)$ on $\delta_2$ and $\delta_D$ is
through their sum. $P(\bar{\nu}_e \to \bar{\nu}_e)$ in Eq.
(\ref{PeeBB}) enjoys having seven oscillatory modes proportional to
1,  $\cos (\Delta_0-\Delta_1)$, $\cos (\Delta_0-\Delta_2)$, $\cos
(\Delta_0-\Delta_3)$, $\cos (\Delta_1-\Delta_2)$, $\cos
(\Delta_1-\Delta_3)$ and $\cos (\Delta_2-\Delta_3)$. In principle,
if the amplitudes of all these modes are measured, all mixing angles
and the two phases entering Eq.~(\ref{PeeBB}) can be extracted. More
precisely, by studying the oscillatory modes of the $\bar\nu_e$
survival probability, the values of $\theta_{12}$, $\theta_{13}$,
$\alpha$, $\beta$, $\gamma$, $\delta_1$ and $\delta_2+\delta_D$ can
be extracted. The only degeneracy will be between $\delta_2$ and
$\delta_D$ which can be broken by a beta-beam or super-beam or
neutrino factor facility which measures $P(\nu_e \to \nu_\mu)$ or
$P(\nu_\mu \to \nu_e)$.
\item
Drawing these figures,  we have taken the ordering of the mass
eigenstates to be normal and have set $\delta_D=0$. As discussed
before, in the case that only one of the mixing angles $\alpha$,
$\beta$ or $\gamma$ is nonzero, the CP-violating phases do not
appear in $P(\bar{\nu}_e \to \bar{\nu}_e)$ so had we set $\delta_D$
equal to some other value, the results would not have changed. The
overall behavior does not also depend on the scheme of mass ordering
({\it i.e.,} normal vs inverted).
\end{itemize}

\section{Conclusions}
We have explored the capacity of the proposed JUNO and RENO-50
setups to probe the SSNS. The mass splitting between the new sterile
neutrino $\nu_0$ and the active neutrino $\nu_1$ is taken of order
of $\Delta m_{01}^2 \sim 10^{-5}~{\rm eV}^2$ and its mixing with
$\nu_1$, $\nu_2$ and $\nu_3$  are respectively denoted by $\alpha$,
$\gamma$ and $\beta$. We have studied the bound that can be imposed
on these mixing angles if their true values vanish (corresponding to
standard three-neutrino mass scheme). We have found that the values
of $\sin^2\alpha$ and $\sin^2\gamma$ down to a few $10^{-3}$ can be
probed by these setups after five years of data taking. With this
setup, it will be possible to marginally probe the parameter space
indicated by the low energy solar neutrino data \cite{Holanda2}. We
have studied the dependence of the sensitivity to baseline and we
have found that the JUNO setup is  close to an ideal setup for the
purpose of probing SSNS. Unlike the case of determining sgn($\Delta
m_{3  1}^2$), farther sources at  distances of order of 200-300 km
are crucial for improving the sensitivity to SSNS. We have shown
that the background and the uncertainty in the flux normalization as
well as the energy resolution are not serious issues for improving
the sensitivity to the SSNS. As a result, by enlarging the detector
size and/or having more numerous and/or powerful sources and/or
prolonging the data taking period, the sensitivity to SSNS can be
increased to a desired level.

 \section*{Acknowledgements}
  The authors would like to thank M.M. Sheikh-Jabbari for careful reading of the
  manuscript. They are also grateful to the anonymous referee for
  useful remarks.
P.B. acknowledges Lashgari for technical help in running the
computer codes.
 Y.F.  acknowledges partial support from the  European Union FP7  ITN INVISIBLES (Marie Curie Actions, PITN- GA-2011- 289442).

%%%%%%%%%%%%%%%%%%%%%%%%%%%%%%%%%%%%%%%%%%%%%%%%%5
%%%%%%%%%%%%%%%%%%%%%%%%%%%%%%%%%%%%%%%%%%%%%%%%%5
%%%%%%%%%%%%%%%%%%%%%%%%%%%%%%%%%%%%%%%%%%%%%%%%%%%%%%%5
%%%%%%%%%%%%%%%%%%%%%%%%%%%%%%%%%%%%%%%%%%%%%%%%%%%%%%%%%

%%%%%%%%%%%%%%%%%%%%%%%%%%%%%%%%%%%%%%%%%%%%%%%%%%%%%%%5
%%%%%%%%%%%%%%%%%%%%%%%%%%%%%%%%%%%%%%%%%%%%%%%%%%%%%%%%%

%
%%%%%%%%%%%%%%%%%%%%%%%%%%%%%%%%%%%

%%%%%%%%%%%%%%%%%%%%%%%%%%%%%%%%%%%

%%%%%%%%%%%%%%%%%%%%%%%%%%%%%%%%%%%

%%%%%%%%%%%%%%%%%%%%%%%%%%%%%%%%%%%
\begin{figure}[t]
\vspace{-5 cm} \hspace{+7mm}
\includegraphics[bb=0 0 1000 1000,scale=0.75]{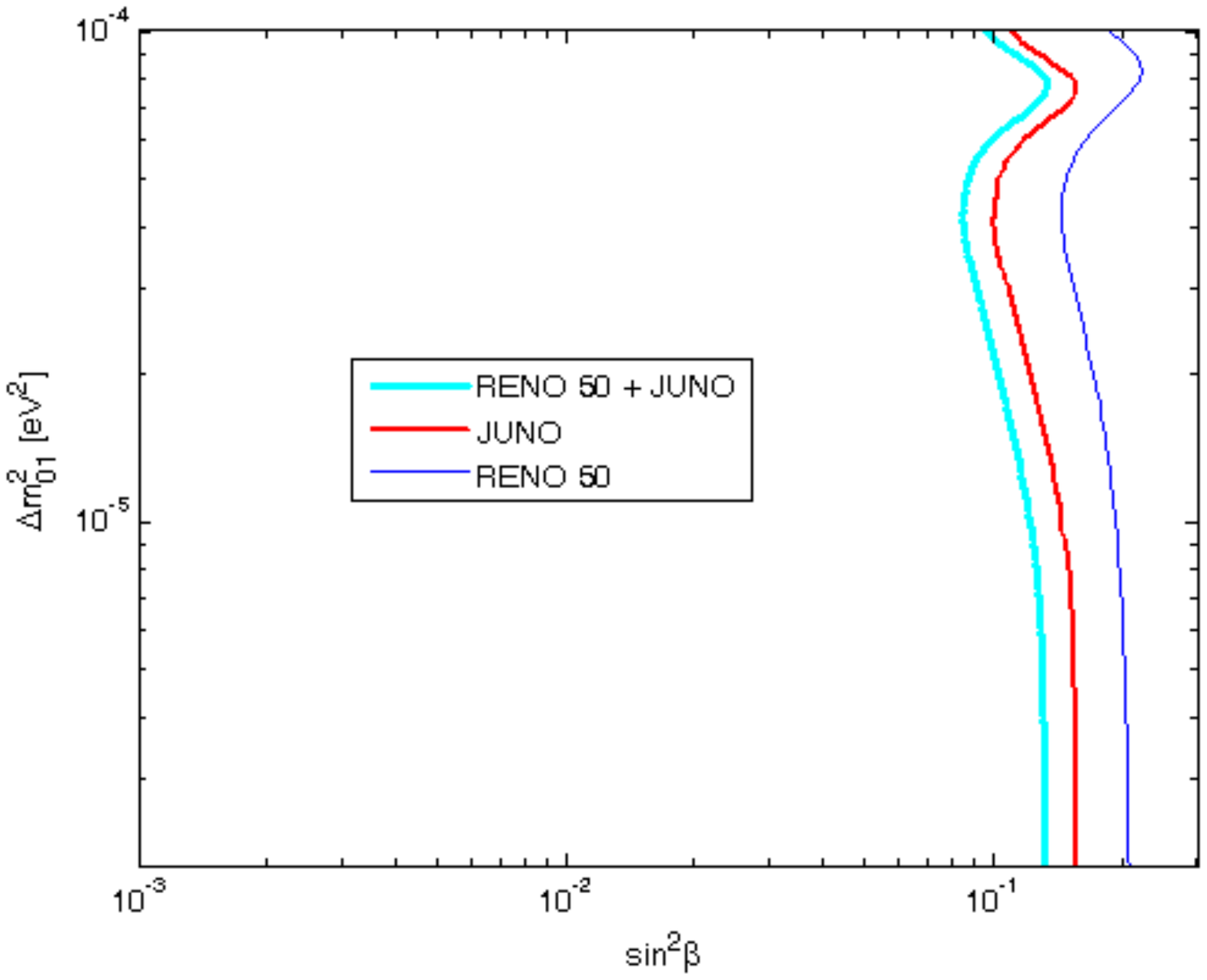}
\caption{The 95\% C.L. upper bound on $\sin^2 \gamma$ versus $\Delta
m_{01}^2$. The rest of the description is similar to that of Fig.
\ref{alpha-bound}. \label{beta-bound}}
\end{figure}
%%%%%%%%%%%%%%%%%%%%%%%%%%%%%%%%%%%
%%%%%%%%%%%%%%%%%%%%%%%%%%%%%%%%%%%
%\begin{figure}[t]
%\vspace{-5 cm} \hspace{+7mm}
%\includegraphics[bb=0 0 1000 1000,scale=0.75]{beta-L.pdf}
%\caption{ The 95\% C.L. upper bound on $\sin^2 \beta$ versus the
%baseline. The rest of the description is similar to that of Fig
%\ref{alpha-L}.\label{beta-L}}
%\end{figure}
\end{document}